\documentclass[epj]{svjour}
\usepackage{graphics}
\usepackage{amsmath}
\usepackage{amsfonts}
\usepackage{color}
\usepackage{epsfig}

\newcommand{\ket}[1]{\left|#1\right\rangle}
\newcommand{\bra}[1]{\left\langle#1\right|}
\newcommand{\da}{^\dagger}

\begin{document}
\title{Light with orbital angular momentum interacting with trapped ions}
\author{Christian Tom\'as Schmiegelow\inst{1} \and Ferdinand Schmidt-Kaler\inst{2}% etc
}                     % Do not remove
\institute{Departamento de Fisica, FCEyN UBA and IFIBA, Conicet, Buenos Aires, Argentina\and
QUANTUM, Institut f\"ur Physik, Staudingerweg 7, Johannes Gutenberg-Universit\"at Mainz, 55099 Mainz, Germany
}
\date{Received: date / Revised version: date}
\abstract{We study the interaction of light beams carrying angular momentum with a single, trapped and well localized ion. We provide a detailed calculation of selection rules and excitation probabilities for quadrupole transitions. The results show the dependencies on the angular momentum and polarization of the laser beam as well as the direction of the quantization magnetic field. In order to optimally observe the specific effects, focusing the angular momentum beam close to the diffraction limit is required. We discuss a protocol for examining experimentally the effects on the S$_{1/2}$ to D$_{5/2}$  transition using a $^{40}$Ca$^+$ ion. Various applications and advantages are expected when using light carrying angular momentum: In quantum information processing, where qubit states of ion crystals are controlled, parasitic light shifts could be avoided as the ion is excited in the dark zone of the beam at zero electric field amplitude. Such interactions also open the door to high dimensional entanglement between light and matter. In spectroscopy one might access transitions which have escaped excitation so far due to vanishing transition dipole moments.}

 % end of PACS codes

\maketitle

\section{Introduction}
\label{intro}

The interaction of light with matter is of fundamental interest, and with the advent of single atom and single ion trapping techniques it became a flourishing field, stimulating many applications including fundamental investigations, high resolution spectroscopy, and quantum information processing~\cite{blatt2008a}. In particular, for the case of interactions with beams with orbital angular momentum (OAM-beam) with atoms various attractive applications have been found: the special properties of the beams may be employed for keeping atoms in hollow-beam dipole traps, the beams may convey angular torque on trapped particles~\cite{friese1996optical} or ensembles of quantum degenerate ultra-cold gases\cite{kulin2001single,rychtarik2004two,andersen2006quantized}, also hollow laser beams have been proven to improve the spatial resolution of fluorescence microscopy in STED techniques~\cite{hell1995ground}. Detectors of electric or magnetic fields in the center of a hollow light beam have been investigated theoretically \cite{KLIMOV}. Most prominently, however, and realized in several seminal works, the OAM single-photon twin beams have proven to  carry quantum information in their different degrees of freedom~\cite{mair2001entanglement,franke2002two}, enlarging the toolbox of quantum entanglement by the higher dimensionality of the Hilbert space.

\begin{figure}
\resizebox{0.45\textwidth}{!}{%
\epsfig{file=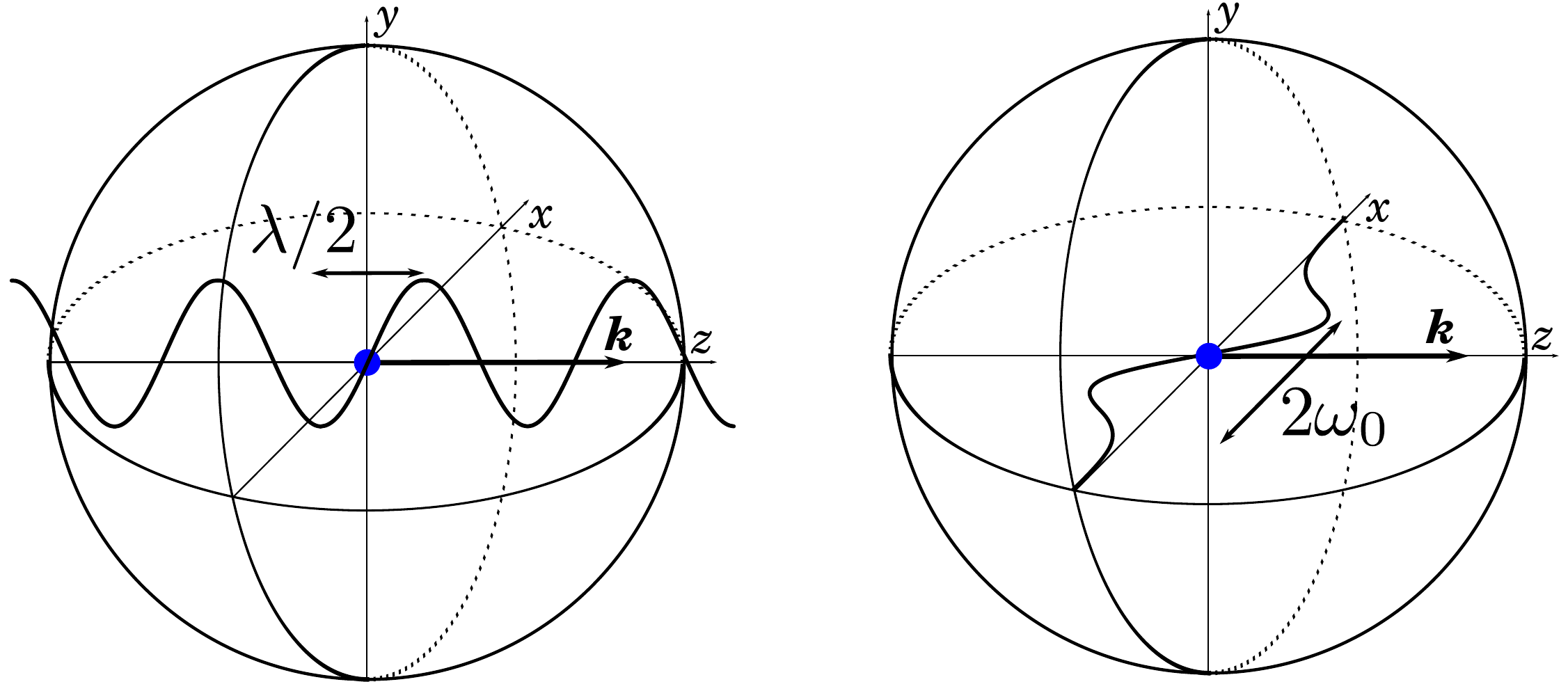, angle=0, width=0.45\textwidth}
}
%\vspace{-0.5cm}       % Give the correct figure height in cm
\caption{Sketch of the relevant electric field gradients in a conventional electric quadrupole transition (left) and a transition with a OAM beam (right). An atom or ion in the center (blue) interacts with a laser beam with wave vector $\mathbf{k}$ along the $z$-axis. For the normal electric quadrupole transition the important field gradient is along the propagation direction and sales with the wavelength $\lambda$. For the OAM-mediated transition the relevant gradient is due to the spatial structure of the beam and it lies in the plane orthogonal to the propagation direction (only projection on the $x$-axis is shown). The transversal gradient for an OAM beam scales inversely with the beam radius $w_0$.}
\label{fig:intuitive}       % Give a unique label
\end{figure}

In this contribution we study the application of laser beams with angular momentum for quantum information processing with trapped ions: In the case of quantum computation, and for implementing a qubit one requires narrow optical transitions between very long lived electronic states. The interaction strength between electromagnetic radiation and matter is described using a multipole expansion of the electro magnetic field~\cite{JACKSON} coupled with the corresponding moments of the atomic charge distribution. While the electric dipole moment couples with the electric field amplitude as \textbf{d} $\cdot$ \textbf{E}, the quadrupole transition moment interacts with next multipole moment of the field, proportional to the gradient  $Q\nabla$\textbf{E}. The use of dipole-forbidden quadrupole transitions is of high interest when long lived qubits states are required. One aims for high gate speeds, in order to overcome various decoherence mechanisms, and to fulfill the criteria for quantum computing~\cite{divincenzo2000physical}. In order to maximize the qubit rotation frequency the quadrupole excitation Rabi frequency $\Omega$ must be increased. To do so one would like to shape the incoming light field in such  a way that the gradient of \textbf{E} is increased. For an incoming plane wave $E \sim \exp(-\emph{i}k_z z)$, and also a weakly focused light beam, the only gradient of the electric field appears along its propagation direction and it magnitude scales proportional to $k \sin(k_z z)$, as depicted in Fig.~\ref{fig:intuitive}a. If, instead of a traveling wave, one had a standing wave, gradient effects are very clearly revealed. In such a standing wave the electric field is $E \sim \sin(k_z z)$ and the gradient becomes maximum in the nodes of the field, and thus the quadrupole excitation strength is optimal at the location where the electric field and also the light intensity becomes zero. This counter-intuitive fact has been demonstrated experimentally with a single ion  which was positioned at different locations inside an optical cavity~\cite{mundt2002coupling}. Here we investigate the excitation of quadrupole qubit transitions with an OAM-laser beam, and discuss its advantages for qubit rotations.
After a calculation of the transition matrix elements in Sect.~\ref{sec:lightion}, we outline a scheme for an experimental realization in Sect.~\ref{sec:exp}. It is of high importance for these proposed experiments, that current Paul trap technology allows to confine and cool a single ion to a spatial wavepacket with an extension of $\leq$ 10~nm, and even to position such a wavepacket  with nm accuracy~\cite{wilson2003,eschner2003sub,POSCH2010}. The discussion of advantages of the qubit excitation at zero light intensity, and possible implications for high resolution spectroscopy of narrow transitions follows in Sect.~\ref{sec:outlook} together with the conclusions of this article.

%Forbidden transitions and the Roos diagrams. -- Stark shift and spectroscopy.
% Idea of the general scheme. Laser induced transitions and rabi flops.
%-- A comment on quantization axes, conserved quantities and, complete sets of commuting observables. Explain the axes!

\section{Ion-Light Interaction}
\label{sec:lightion}
\begin{figure}[b]
\begin{center}
\resizebox{0.35\textwidth}{!}{%
\epsfig{file=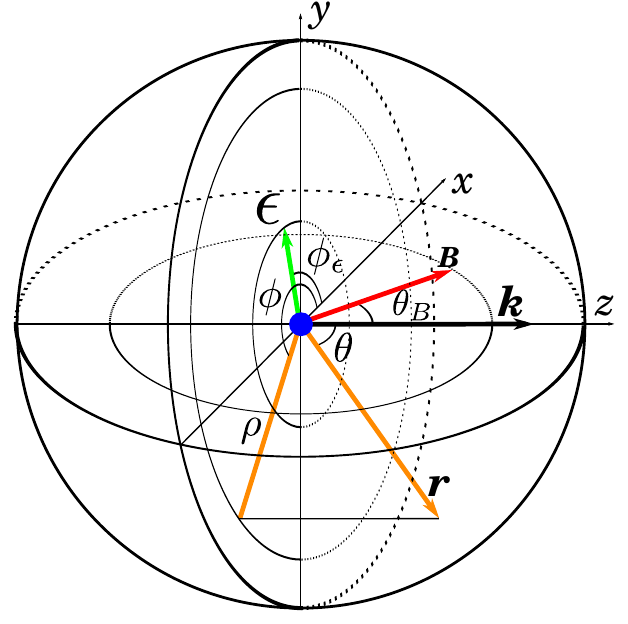, angle=0, width=0.35\textwidth}}
\end{center}
\caption{Labelling of the vectors and quantities involved in the interaction. The laser has wave vector $\mathbf{k}$ along the $z$-axis and has linear polarization making an angle $\phi_\epsilon$ with the $x$-axis. The magnetic field lies in the $x$-$z$ plane making an angle $\theta_B$ with the $z$-axis. The integration variables $\mathbf{r}$ and $\rho$ are characterized by the polar angles $\theta$ and $\phi$.}
\label{fig:axes}       % Give a unique label
\end{figure}

As a general scheme we consider a single atom or ion which interacts with a light field described by a vector potential $\mathbf{A}$. As depicted in Fig.~\ref{fig:axes} the beam travels in the $z$ direction with linear polarization forming an angle $\phi_\epsilon$ with the $x$ axis, the atom is aligned with respect to a quantization magnetic field $\mathbf{B_0}$ which lies in the $x$-$z$ plane forming an angle $\theta_B$ with the $z$ axis. We will consider the atom or ion in an initial state $\ket{i}$ making a transition to a final one $\ket{f}$. The relevant matrix elements characterizing the transition are then:
\begin{equation}
\bra{f}\mathcal{H_I}\ket{i}
\propto \bra{f}\mathbf{A}\cdot\mathbf{p}+\mathbf{p}\cdot\mathbf{A}\ket{i}
\propto \omega_{fi}\bra{f}\mathbf{r}\cdot\mathbf{A}\ket{i}
\label{eq:interaction}
\end{equation}

We consider the most general vector potential for a paraxial beam with orbital angular momentum: a Laguerre Gaussian beam. The vector potential for such a beam at the focus $w_0$ is written in cylindrical coordinates ($\rho$, $\phi$, $z$) as~\cite{allen1992orbital}:
\begin{eqnarray}
\mathbf{A}_{lp}&=&\mathbf{A_0}
\underbrace{
	\sqrt{\frac{2 p!}{\pi(|l|+p)!}}
	\left(\frac{\sqrt{2}\rho}{w(z)}\right)^{|l|}
	\mathcal{L}_p^{|l|}\left(\frac{2\rho^2}{w^2(z)}\right)
	\exp(il\phi)
}_{Orbital\; Angular\; Momentum}\nonumber\\
&&\times\underbrace{	
	\exp(ikz)
	 \frac{w_0}{w(z)}\exp\left(\frac{-\rho^2}{w(z)}+\frac{ik\rho^2}{2R(z)}+i\Phi_g(z)\right).
}_{Gaussian\;	 Beam}
\end{eqnarray}
Where $w(z)$, $R(z)$ and $\Phi_g(z)$ are the usual radius, curvature and Gouy phase of a paraxial beam and $\mathbf{A}_{lp}$ is a quantity proportional to the amplitude of the field at the center of the focus~\cite{allen1992orbital,saleh1991fundamentals}. The index $l\in\mathbb{Z}$ corresponds to the angular momentum while the index $p\in\mathbb{N}_0$ indicates the amount of radial nodes. The functions $\mathcal{L}_p^{|l|}$ are the generalized Laguerre polynomials. The following subsections first review the calculation of the matrix elements for the standard electric quadrupole transition with beams with Gaussian transversal shape ($\mathbf{A}_{00}$) and then extend the analysis to orbital angular momentum mediated transitions. In particular we study the case of a beam with total angular momentum 1$\hbar$ ($\mathbf{A}_{10}$ or OAM=1). Finally we discuss their differences. Details on the analytic calculation are given in Appendix \ref{app:calc}.

\begin{figure*}
\epsfig{file=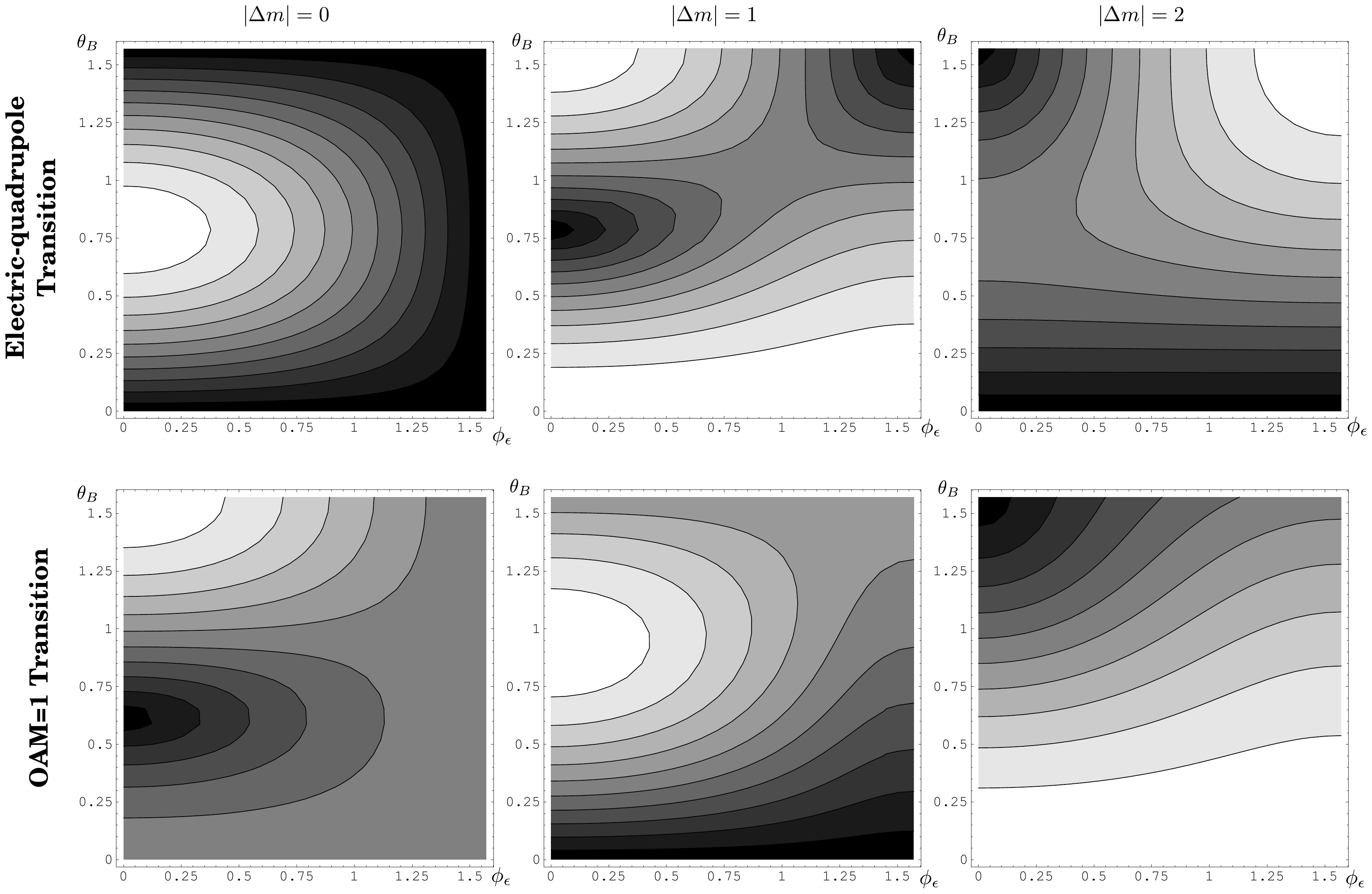, angle=0, width=0.9\textwidth}
\caption{Contour plots of the transition matrix elements for both the standard electric quadrupole transition (top row) and a transition in which a OAM=1 beam (bottom row). Each column is for a transition with different change in the angular momentum projection $\Delta m$ in the direction of the magnetic field. Results are shown as a function of the angle of the linear polarization of the beam $\phi_\epsilon$ and the direction of the magnetic field $\theta_B$. Values range from 1 (white) to 0 (black) in steps of 0.1; additionally a pre-factor has to multiplied for each plot (see Table \ref{tab:elems}).}
\label{fig:roos}       % Give a unique label
\end{figure*}	

\subsection{Review of the standard electric quadrupole excitation}
We restrict ourselves to the case where the atom or ion is localized in the plane of the focus of the beam where $\omega(z)=\omega_0$. Under reasonable and usual experimental working conditions the extension of the the atom's or ion's wavepacket $a_0$ is much smaller than the wavelength of the interacting light: $\lambda\gg a_0$. Then, for the calculation of the matrix elements one can do the approximation $w_0\gg\rho$ since outside that region the integrand in nearly null. This is discussed in many textbooks in the framework of electric dipole and quadrupole approximation for a Gaussian beam  or plane wave~\cite{sakurai1985modern}. To do so, one expands the vector potential of the Gaussian beam $\mathbf{A}_{00}$ in a series in $k$.
\begin{equation}
\mathbf{A}_{00}\approx\mathbf{A_0}(1+ikz).\label{eq:oam0}
\end{equation}
The $\mathbf{A_0}$ term is responsible for the well known dipole-allowed transitions since in the transition matrix \ref{eq:interaction} it gives terms proportional to $r$, that is, a spherical tensor of rank one. The following term,  $\mathbf{A_{0}}ikz$, will allow for quadrupole transitions, where more than one unit of angular momentum per photon is exchanged. As is detailed in the Appendix, this results from the fact that this interaction term can be decomposed into a sum of spherical harmonics up to rank two. However, this effect scales as $\mathbf{A_0}ka_0$,  and is normally very small. For optical transitions in atoms and ions we find typical values of $ka_0\approx10^{-4}$. By choosing the right experimental parameters one can still observe such weak transitions: one may increase the power of the interacting laser or focus down to a small waist size thus increasing  $\mathbf{A_0}$. If the desired transition is driven between two long lived states, the Rabi frequency may overcome the linewidth, leading to coherent oscillations of the upper state population. Typical examples of focused beams for the excitation of ions have been reported in~\cite{eschner2003sub,nagerl1999laser,guthohrlein2001single} where Rabi frequencies of MHz are observed for ions in a few mW laser beam with waist size of 1.5$\mu$m. The selection rules for the S$_{1/2}$ to D$_{5/2}$ transition have been evaluated~\cite{roos2000controlling} and depend on the Zeeman sub-level of the ground and the excited state, the direction and polarization of the laser beam with respect to the magnetic quantization axis, see Fig.~\ref{fig:axes}.

\subsection{Electric quadrupole excitation in a OAM-beam}
For a beams with OAM the same approximations as before lead to different terms in the multipole expansion of the beam. In particular for a beam with 1$\hbar$ of angular momentum (OAM=1) the leading term of the vector potential $\mathbf{A}_{10}$ is:
\begin{equation}
\mathbf{A}_{10}\approx\mathbf{A_0}\frac{\sqrt{2}\rho}{w_0}e^{i\phi}\label{eq:oam1}
\end{equation}
With this kind of interaction, dipole-forbidden transitions are allowed. This is because the whole interaction term can be decomposed as a sum of spherical tensors of rank up to two. That means: terms with quadratic dependence in the coordinates. It may be interesting to note that this term appears as the leading term in the expansion and not as a higher order correction as in the case of the fundamental Gaussian beam. However the interaction due to the OAM-beam does not exhibit a strength comparable with that of dipole interactions. This is due to the fact that in this type of interaction, the radius $w_0$ of the beam at the focal plane appears as a relevant scaling parameter. It is also interesting that for OAM-beams the interaction strength is not directly related to the wavelength as for the electric quadrupole transition in a Gaussian. In contrast, for this case, the limit is given by how tight one can focus the beam, i.e. to the diffraction limit.

\subsection{Discussion of differences between fundamental Gaussian and OAM=1 beam}
Looking at expressions \ref{eq:oam0} and \ref{eq:oam1}, one sees that in order for the effects of the OAM-mediated transition to be of the order of the electric quadrupole the focusing must be such that $w_0=\lambda/\sqrt{2}\pi$. That is just $\sqrt2$ over the diffraction limit. Such focusing at the diffraction limit will show OAM effects only $\sqrt2$ smaller than electric quadrupole ones, enough to be clearly observable. The calculation of the exact transition probabilities $\bra{f}\mathbf{r}\cdot\mathbf{A_0}kz\ket{i}$ and $\bra{f}\mathbf{r}\cdot\mathbf{A_0}\frac{\sqrt{2}\rho}{w_0}e^{i\phi}\ket{i}$ can be carried out analytically. The general procedure is outlined in Appendix \ref{app:calc}. It is an extension of the method presented by D. James~\cite{james1998quantum} and later used for the S$_{1/2}$-D$_{5/2}$ transition in calcium ions by Ch. Roos~\cite{roos2000controlling}. Fig.~\ref{fig:roos} shows transition probabilities for such a $\Delta l= 2$ electric dipole forbidden transition for $|\Delta m|=0,1,2$ and all possible linear polarizations $\phi_\epsilon$ and magnetic field directions $\theta_B$.
	
The behaviour for $\theta_B=0$ can be clearly understood in qualitative terms. As the magnetic field is aligned with the $k$ vector of the incident light field, the quantization axes set by these both are the identical~\cite{enk1994selection,picon2010photoionization}. This implies a rotation symmetry around this axis which manifests itself as a conservation law: the conservation of $J_z$, the angular momentum in the $z$ direction. It is clearly seen in Fig.~\ref{fig:roos} that for $\theta_B=0$ in the electric quadrupole case  only  the $|\Delta m=1|$ transition is allowed and there is no dependence on the orientation of the polarization. This is reasonable since photons from a Gaussian beam have angular momentum projection in this direction $S_z=\pm1$. For OAM=1 beams the allowed transitions are $|\Delta m=2|$ and $|\Delta m=0|$. These two possibilities are for the photon's spin aligned parallel or anti-parallel to the orbital angular momentum of the photon adding up to a total $J_z$ transfer from the photon to the atom or ion of 2 or 0.

The behaviour for non collinear configuration $\theta_B\neq0$ is more complex and we cannot give an explanation in qualitative terms. In this case $J_z$ is not a good quantum number. Different combinations of polarization and magnetic field angles make all possible transitions accessible. In fact, for the standard electric quadrupole interaction, this has been used  for precision measurements~\cite{kreuter2005experimental}. By choosing, for example, $\theta_B=\pi/2$ the $|\Delta m=1|$ or $|\Delta m=2|$ transition can be selected just by rotating the polarization.

One final remark about conservation laws in S-D transitions. Consider the case where the atom or ion gains two units of total orbital angular $\Delta l =2$ and this occurs in the regular electric quadrupole interaction where the beam has no extra orbital angular momentum. Where does the extra unit of angular momentum come from? The angular momentum comes from the gradient of the field in the $z$ direction, the same way the gradient of the field in the transverse plane is responsible angular momentum in a Laguerre Gaussian beam. It is a common misconception to think that ideal plane waves only have intrinsic angular momentum associated with the photon's polarization. Any gradient in the field, even the one in the propagation direction, can account for extra angular momentum.

\subsection{Very strong focusing}
It is important to state that very strong focusing to the diffraction limit leads to unexpected effects~\cite{dorn2003focus,monteiro2009angular,jauregui2004rotational}, where orbital angular momentum and polarization can no longer treated independently. With such strong focusing one would still be able to excite dipole forbidden transitions, but they will not necessarily obey the transition rules and angular dependencies calculated here. \footnote{Both, the calculation of the quadupole excitation in very focused beams beyond paraxial approximation and the calculation for a atom not positioned in the plane of the focus are future work.}  To observe pure OAM effects, as described by these calculations and detailed below, it is best not to focus the beam down to the diffraction limit.

\section{Proposed experimental scheme}
\label{sec:exp}
In this section we analyse focusing requirements and give experimental parameters for observing such an effect in a $^{40}$Ca$^+$ ion trap. In particular we study the dipole forbidden S$_{1/2}$ to D$_{5/2}$ transition at 729~nm. We will show that under appropriate conditions it is possible to observe OAM mediated transitions and clearly distinguish them from regular electric quadrupole transitions in a fundamental Gaussian mode.

\subsection{Detection of OAM beam excitation using Rabi oscillations}
The quantitative test of the theoretical considerations may be done with a single trapped ion and is based on the electron shelving technique, in a sequence of the following steps: (i) the ion is laser cooled on a dipole allowed transition S$_{1/2}$ to P$_{1/2}$ near 397~nm, optical pumping into the metastable states D$_{5/2}$ and D$_{3/2}$ is excluded by radiation near 866~nm and 854~nm. (ii) The ion is optically pumped with $\sigma$+ radiation near 397~nm into the Zeeman ground state S$_{1/2}$, m=+1/2. (iii) A laser pulse, tuned to a chosen transition of the S$_{1/2}$ to D$_{5/2}$ near 729~nm, with well defined length and intensity is sent to the ion. (iv) The excited state probability is detected as the ion is illuminated with radiation near 397~nm and 866~nm, and the laser induced fluorescence is detected. A high level of fluorescence means that the ion is measured in the S$_{1/2}$, while no fluorescence is emitted if the excitation to the D$_{5/2}$ was successful. This sequence is repeated and the average over hundreds of experiments reveals the mean excitation probability. Control parameters of the quadrupole excitation step (iii) are varied, especially the time duration, to directly measure the Rabi frequency.

\begin{figure}
\begin{center}
\epsfig{file=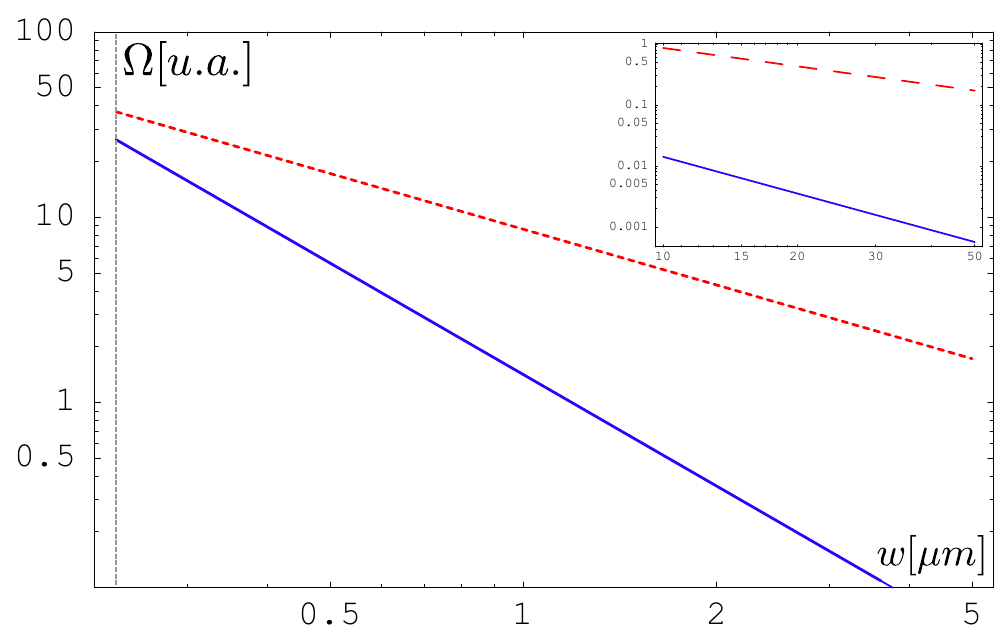, angle=0, width=0.45\textwidth}
\end{center}
%\vspace*{0.5cm}       % Give the correct figure height in cm
\caption{Rabi frequency $\Omega$ (in arbitrary units) of the transition as a function of focus size $w$ for beam with a wavelength of 729~nm for the electric quadrupole transition (red dashed) and the transition with orbital angular momentum 1 (blue line). The vertical line indicates the diffraction limit. Inset: Same for weakly focused beams.}
\label{fig:focusing}       % Give a unique label
\end{figure}

As stated in the previous chapter, if the beam is focused almost to the diffraction limit, the ratio between the standard electric quadrupole Rabi frequency and that one mediated by the OAM=1 transition approaches $\sqrt{2}$. If the beam is less strongly focused then the relative intensity of the electric quadrupole transition will be correspondingly stronger. Fig.~\ref{fig:focusing} shows the intensities of these two transitions as a function of focusing for two different regimes: strong and weak focusing.  It is seen that even if the focus size is 10 times larger than the diffraction limit, the relative intensity of both transitions is just a little larger than 10. This is a ratio that can be readily resolved with current coherent spectroscopy techniques. For weaker focuses,  $>15\mu m$ beam waist, the ratio grows to about 100, making the OAM-excitation harder to observe. The relative ratio between the two transitions is $\sqrt{2}\pi w / \lambda$. The calculated values plotted in Fig.~\ref{fig:focusing} take the explicit dependence of $A_{lp}$ on the beam waist into account. For the electric quadrupole and the OAM=1 transitions the intensities scale as $k/w$ and $\sqrt{2}/w^2$ respectively.

The above considerations are valid for a situation when both transitions, with the regular quadrupole and the OAM, are probed for a polarization and magnetic field angles such that the transition to a given level is maximized. As can be seen from Fig.~\ref{fig:roos} there is no such configuration where this condition can be fulfilled for one setting of $\theta$ and $\phi$. One might even find regions where the transition is exclusively from the OAM effect. For example, by choosing the collinear configuration $\theta_B=0$ different transitions will be allowed depending if the beam is purely Gaussian or has one unit of angular momentum. In such a case, Fig.~\ref{fig:roos} shows that the transition $|\Delta m|=1$ is exclusively allowed for a purely Gaussian beam while the transitions $|\Delta m|=2$ and $\Delta m=0$ are allowed for an OAM=1 beam. Transitions to different Zeeman levels can be clearly resolved in $^{40}$Ca$^+$ as they split by a couple of MHz already in a quantization magnetic field of a few Gauss.

\begin{figure}
\begin{center}
\epsfig{file=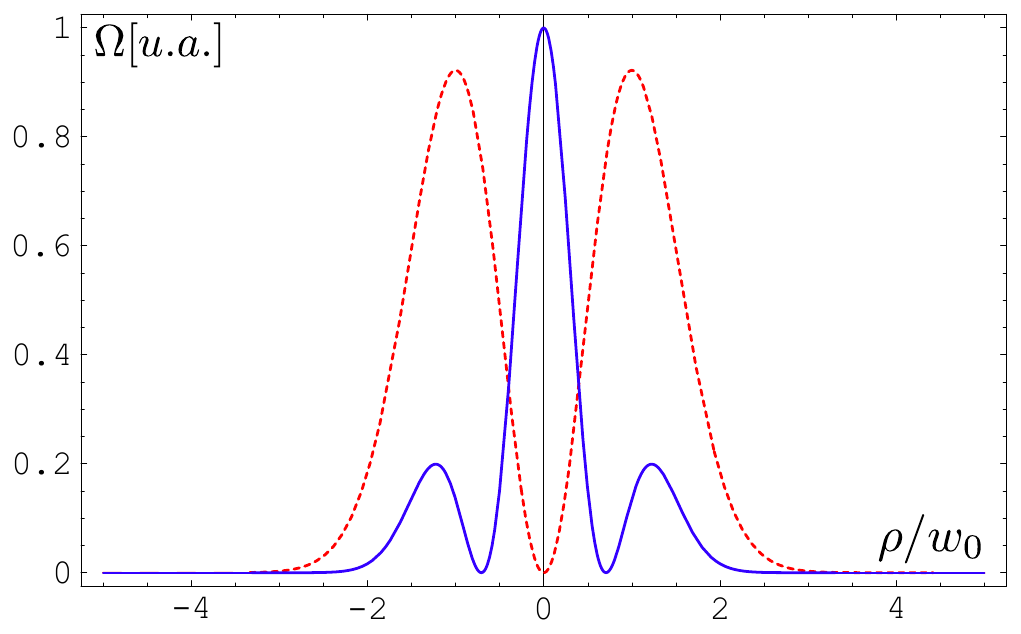, angle=0, width=0.45\textwidth}
\end{center}
%\vspace*{0.5cm}       % Give the correct figure height in cm
\caption{Normalized transverse profile of the Rabi frequency for a OAM=1 beam. When the interaction is due a $\mathbf{A_{0}}ikz$ term then, $\Omega$ it is proportional to the intensity of the beam revealing its doughnut shape (red dashed line). In contrast, the OAM-mediated transition the Rabi frequency is proportional to the absolute value of the gradient of the field showing a central peak and two smaller peaks at the sides (blue line).}
\label{fig:profiles}       % Give a unique label
\end{figure}

Another possible test for these kind of transitions would by a sub micrometer positioning of ions with respect to the beam. Positioning resolutions of $\leq 1\mu m$ can be achieved in segmented traps by designing appropriate trapping potentials~\cite{mundt2002coupling,guthohrlein2001single,HUBER}. If subject to a OAM-transition, an ion moved across a OAM=1 beam should exhibit a interaction profile with a big peak at the center of the beam and two smaller peaks at each side. However, for this same beam, if the interaction mediated by the term $\mathbf{A_{0}}ikz$ of the expansion of a Gaussian beam, one should only see the beam intensity profile: two equal peaks at each side of the focus. Both these behaviours are sketched qualitatively in Fig.~\ref{fig:profiles}.

\subsection{Specifics and advantages of OAM-beam excitation}
The experimental observation of OAM-beam excitation is an open task, but we would discuss here some peculiarities, especially as they would lead to unique, possibly important effects:

One aspect, not discussed so far is the \emph{transfer of momentum} from the laser light to the motion of trapped ions. Such change of vibrational quantum states is important for laser cooling~\cite{eschner2003laser,wineland1998experimental}, quantum state engineering, and quantum gate operations. The focussing of the OAM-beam and the laser excitation on the quadrupole transition requires a newly defined Lamb Dicke factor.

The regular Lamb Dicke factor relates the extension of the ion wavepacket in its ground state with the wavelength of the light field, with $ \eta = k \Delta x_0 $ where \emph{k} denotes the projection of the k-vector on the oscillation direction of the specific mode\footnote{One can take the projection of $\textbf{k}$ in direction of a specific mode into account by a prefactor of sin($\alpha$)}. Thus, the Lamb Dicke factor $ \eta $ scales with $ k=(2\pi)/\lambda $~\cite{wineland1998experimental}. The ground state wavepacket extension reads as $\Delta x_0 = \sqrt{\hbar/2m\omega}$. The Lamb Dicke regime is defined also for thermally excited ion wave packets by replacing $\Delta x_0$ with $\Delta x_n = \Delta x_0 \sqrt{\overline{n}+1}$, where $\overline{n}$ is the mean thermal quantum number. For a narrow transition, the  so called strong confinement regime holds when the width of the transition $2\pi \Gamma$ is well below the vibrational frequency $\omega_{trap}$. In this limit, the laser can be tuned to either the carrier transition between Fock states $|n\rangle \rightarrow |n\rangle$ or to the sideband transitions where the vibrational quantum number undergoes a change to $|n\pm1\rangle$.

For the OAM-excitation, the laser waist $w$ replaces $\lambda/2$ to \emph{define a new Lamb Dicke parameter}, which reads now as $\mu=(\pi/w_0)\sqrt{\hbar/2m\omega} $. If an OAM=1 beam is employed, the electrical field depends as E($\rho) \sim \rho $ (eq u.~\ref{eq:oam0}) on the transversal coordinates, the derivative of this field dependency leads to a constant gradient. Thus, only the carrier transitions between identical Fock states are excited\footnote{Or transitions between states with identical parity, thus the second sideband to $|n\rangle \rightarrow |n\pm2\rangle$}. Contrary, for a OAM=2 beam the electric field scales quadratically with E($\rho) \sim \rho^2 $ (Eq.~\ref{eq:oam0}) such that the gradient becomes a linear function of $\rho$. Here, only and only sideband transitions from $|n\rangle \rightarrow |n+1\rangle$ or $|n\rangle \rightarrow |n-1\rangle$ exhibit non-zero transition matrix elements! The situation resembles to that  one in a  Gaussian standing wave, where an ion at the node will undergo only carrier transitions and in the anti-nodes only first sideband transitions~\cite{mundt2002coupling}\footnote{The gradient of the electric field is in the direction of the beam propagation in the Gaussian case and perpendicular to the beam propagation in the OAM case}. Consequently, one may adapt improved cooling schemes with such laser beams that avoid disturbing carrier excitation and lead to improved cooling limits~\cite{MORIGI,CIRAC}. Furthermore, the quantum gate fidelity may be improved if parasitic carrier excitation~\cite{steane2000speed} can be fully excluded due to selection rules. One might even conceive the dynamic modification of the laser beam from a OAM=1 and OAM=2 beam by a spatial light modulator~\cite{grier2003revolution,gibson2004free} during a quantum algorithm, adapted for optimum fidelity of the sequence of gates.

The specific advantage of a quadrupole excitation on the sidebands only without parasitic carrier in the OAM=2 beam is highlighted in the Moelmer Soerenson scheme for \emph{generating entangled ions}~\cite{sorensen1999quantum}. Here, a bichromatic light field with frequency components $\nu = \pm \nu_{trap} + \delta$ is applied such that the initial quantum state $|g_1,g_2\rangle$ evolves into a Bell state $1/\sqrt{2} (|g_1,e_2\rangle + |e_1,g_2\rangle)$. If a high fidelity output state~\cite{benhelm2008} is required one has to avoid carrier excitations which would populate the states $|g_1,e_2\rangle$ and $|e_1,g_2\rangle$. Especially if one aims to quickly reach the Bell state, driving the interaction faster with high laser power, the fidelity of the entanglement operation is reduced. Even careful pulse shaping~\cite{ROOS2008} can not fully avoid a spectral overlap with the carrier resonance where the Rabi frequency is stronger by a factor of $\eta^{-1}$, but is only detuned by about one trap frequency $\nu_{trap}$. For the case of a OAM=2 beam with a linear two-ion crystal aligned along the propagation direction of the beam, both ions could be placed in the center of the light beam. The beam containing the bichromatic field would lead to sideband dynamics via a radial common vibration mode with the modified Lamb Dicke factor $\mu$. Even though $\mu$ would be smaller as compared to $\eta$ (see Fig.~\ref{fig:focusing}) the laser power may be increased and this would largely overcompensate the effect so that gate operation time may be improved.

The excitation of the qubit transition occurs in the fully dark region of the OAM-laser beam, and consequently no dipole transitions are exited. This feature may be helpful to \emph{avoid parasitic AC-Stark shifts}, that require very careful compensation in high resolution spectroscopy and for quantum gate operations~\cite{FSK2004}. The magnitude of such shift scales with $E^2/\Delta$ where $\Delta$ denotes the frequency detuning of the qubit transition from a dipole transition, for example the S$_{1/2}$ to P$_{1/2}$ transition near 397~nm. The finite wavefunction of the ion overlaps to a very small fraction with the focused OAM-beam electric field and the estimation of the integral $(1/\Delta) \int E^2(\rho) |\Psi(\rho)|^2 d\rho$ results in a reduction of the AC-Stark effect to less than 0.1~$\%$ for a 15~nm ion wavepacket inside a 2~$\mu$m OAM=1 beam waist. The implications for high resolution spectroscopy are evident since a narrow, weak line might be investigated without parasitic excitation and frequency shift due to a near-lying dipole transition.

For single qubit rotations in an elementary quantum processor, a Gaussian laser beam is tightly focused to one ion of the ions in a linear crystal and one aims to avoid parasitic excitation of neighboring ions since that affects the gate fidelity. Such addressing~\cite{nagerl1999laser} was reached to a focus size of about 1.5~$\mu$m and has been utterly important to generate  entangled quantum states~\cite{schmidt2003realization}. The OAM=1 beam would allow an \emph{improved addressing}, as the specific OAM-transition, see Fig.~\ref{fig:roos}b can be chosen with high strength, while the neighboring ions do not sense a transversal gradient in the electric field. They might be still touched by a tail of the beam but a clever chosen set of parameters $\theta$ and $\phi$ can be used to exclude any excitation due to the selection rules of OAM$\neq$1 transitions. For example with $\theta$ = $\pi/2$ and $\phi$ =0 the OAM=1 amplitude is maximum on the $\Delta$m=0 transition while the regular Gaussian beam excitations fully vanishes.

\section{Summary and Outlook}
\label{sec:outlook}
We presented the calculation of the quadrupole excitation for a well localized atom or ion in the center of a laser beam with angular momentum. We have also proposed an experimental realization of such interaction where the orbital angular momentum content of photons dominates the light-matter interaction. One of the most amazing features is that the excitation takes place in a region of space near where the light intensity is zero. This also happens with standing waves, but in the case of OAM beams there is no need for optical cavities. Many different applications of this unique feature were sketched: the suppression of parasitic light shifts might be the most important one in the field of quantum information and metrology experiments with single atoms or ions.

Similar OAM dependent transitions should be also accessible in artificial atoms. Quantum dots or and other nano-fabricated quantum systems exhibit selection rules on their transitions which depend on angular momentum~\cite{quinteiro2009electronic,quinteiro2010electronic}. They should all be accessible and controlled by use of photons with OAM. This should allow the generation of high dimensional entanglement between these and other similar devices using photons.

In the future one might extend the method from OAM-beams to evanescent light fields, which could provide even higher electric gradients. Ultra-thin fibres coupled to atoms around \cite{DAWKINS2010,sague2007cold} or microscopic whispering gallery resonators~\cite{louyer2005tunable} would provide an ideal experimental platform for such investigations. While the experiments so far have been concentrating on dipolar coupling matrix elements only, our study could be extended for quadrupole transitions between long lived states. For reaching the tightest focussing even well outside the paraxial approximation with $w \sim \lambda$ one could also employ microwave fields, which may be strongly coupled to Rydberg atoms. This would be of interest for cavity QED with neutral Rydberg atoms \cite{HAROCHE}, or possibly in novel experimental approaches with trapped ions, that are excited to the Rydberg states~\cite{FSK2010}, would open new opportunities to focus strongly radiation onto well localized ions or atoms, positioned in the microwave field. We note that neighboring Rydberg levels of ions near a principle quantum number n=60 are coupled by microwave radiation at cm-wavelengths. Such fields may be readily formed with antennas or lenses, both in the far and the near field.

All experimental preconditions and parameters have been chosen to match to typical ion trap experiments. Thus, the proposed experimental sequence and the detection of OAM=1 transitions in the dark center of the light mode are well in reach with current technology.

Acknowledgements: FSK thanks the VW-foundation and the EU-commission within the FP7-IP AQUTE for financial support.  We thank for helpful discussions with Rene Gerritsma, Frank Ziesel, Max Hettrich, Cecilia Cormick, Ulrich Poschinger and Christian Roos for comments. We acknowledge important contributions in view of a future experimental realization of the scheme by Andreas Walther and Ulrich Poschinger.

%Our study may open the opportunity for microscopy well below the Abbe diffraction limit using OAM=1 beams. The STED technique relies on the dark spot of the beam, where the
%
%Also, high dimension entanglement between angular momentum states of photons and Zeeman levels could be accessible by this method.
%
%\comment{Should we include here something on:-- STED microscopy. sub wavelength imaging, possible also for temperature measurements, but now on the diploe transition, either 866 or 397nm, to what extend does the AC stark effect really vanishes? Overlap of thermal wavepacket with OAM beam E field.
%Solid state single photon emitters.-- Microwave photons.

\appendix
\section{Calculation of selection rules}
\label{app:calc}

\begin{table*}\begin{center}
\begin{tabular}{c|cc}
& Electric Quadrupole & OAM =1\\\hline\noalign{\smallskip}
$\Delta m=0$&
$\frac{3}{\sqrt{10}}| \cos(\phi_\epsilon) \sin (2\theta_B)|$&
$\frac{1}{2\sqrt{10}}|(3\cos(2\theta_B)-1)\cos(\phi_\epsilon)-2i\sin(\phi_\epsilon)|$\\\noalign{\smallskip}
$|\Delta m|=1$&
$\sqrt\frac{2}{5} \;|\cos(\phi_\epsilon) \cos (2\theta_B)+ i  \sin(\phi_\epsilon) \cos (\theta_B)|$&
$\frac{1}{\sqrt{10}} \;|(\cos(\phi_\epsilon)+2\cos(\phi_\epsilon) \cos (\theta_B) + i \sin(\phi_\epsilon)) \sin (\theta_B)|$\\\noalign{\smallskip}
$|\Delta m|=2$&
$\frac{1}{\sqrt{5}} \;|(\cos(\phi_\epsilon) \cos (\theta_B)+ i \sin(\phi_\epsilon)) \sin (\theta_B)|$&
$\frac{1}{\sqrt{5}} \;|(\cos(\phi_\epsilon) \cos (\theta_B)+ i \sin(\phi_\epsilon)) \cos^2 (\theta_B/2)|$
\end{tabular}
\end{center}
\caption{Transition matrix elements for the electric quadrupole and orbital angular momentum transition for different change in the angular momentum projection $\Delta m$ in the direction of the magnetic field. }\label{tab:elems}
\end{table*}

In this appendix the general procedure for the calculation of the transition probabilities is outlined. First the general scheme is presented and then it is illustrated with the detailed calculation for the electric quadrupole and OAM=1 interactions.

In general, for any interaction term, the selection rules can be calculated analytically following a couple of operative steps:
\begin{enumerate}
\item Write the interaction Hamiltonian in terms of products of spherical tensors.
\item Rewrite this product as a sum of spherical tensors using the formula:\begin{equation} \label{eq:prod-sum}
T_{m_1}^{(k_1)}T_{m_2}^{(k_2)}=\sum_{k,m}T_{m}^{(k)} \langle k_1 k_2;km|k_1 m_1;k_2 m_2\rangle.
\end{equation}
Where $\langle k_1 k_2;km|k_1 m_1;k_2 m_2\rangle$ are the usual Clebsh-Gordan coefficients.
\item Rotate this operator an angle $\theta_B$ around the $y$ axis so that they in the same base as the electronic states (which are quantized along the magnetic field). To do so use the Wigner rotation matrices $d_{mm'}^{(k)}$ so that each spherical tensor transforms as
\begin{equation}D\da_y(\theta)T_m^{(k)}D_y(\theta)=\sum_{m'}T_{m'}^{k}d_{mm'}^{(k)*}(\theta).\end{equation}
\item Use the Wigner-Eckart theorem
\begin{equation}
\langle jm|T^{(k)}_q|j'm'\rangle =\langle j||T^k||j'\rangle \langle j'm';kq|jm \rangle
\end{equation}
 to calculate the relative weight of each term in the total transition probability.
\end{enumerate}

We now show the details for the case of a $\mathbf{A_{0}}ikz$ and a OAM=1 interaction. The relevant interaction terms are, respectively:
\begin{eqnarray}
&&\bra{f}
\left[ \frac{x+iy}{\sqrt{2}}\epsilon_r+ \frac{x-iy}{\sqrt{2}}\epsilon_l \right] z\ket{i}\; A_0 k
\\
&&\bra{f}
\left[ \frac{x+iy}{\sqrt{2}}\epsilon_r+ \frac{x-iy}{\sqrt{2}}\epsilon_l \right] r \sin{\theta}e^{i\phi}\ket{i}\; A_0 \frac{\sqrt{2}}{w_0}.
\end{eqnarray}
First (1.) one rewrites the interaction matrices in terms of spherical tensors by using that
\begin{eqnarray}
\frac{x\pm iy}{\sqrt{2}}&=&r\sqrt{\frac{4\pi}{3}} T_{\pm1}^{(1)}\\
z&=&r\sqrt{\frac{4\pi}{3}} T_{0}^{(1)}\\
 \sin{\theta}e^{i\phi}&=&-\sqrt{\frac{8\pi}{3}} T_{-1}^{(1)}.
\end{eqnarray}
These relations are the same as the ones that arise between coordinates and spherical harmonics. The usual labeling of spherical tensors, however inverts the top and bottom indices with respect to spherical harmonics.

Putting all this together we rewrite the two matrix elements as:
\begin{eqnarray}
&&\bra{f}
\left[- T_{1}^{(1)}\epsilon_r+ T_{-1}^{(1)}\epsilon_l \right] T_{0}^{(1)}r^2 \ket{i}\;  A_0 k\frac{4\pi}{3}
\\
&&\bra{f}
\left[-T_{1}^{(1)}\epsilon_r+  T_{-1}^{(1)}\epsilon_l \right]   T_{-1}^{(1)}r^2 \ket{i}\;  A_0 \frac{(-1)}{w_0}\frac{8\pi}{3}
\end{eqnarray}
 Next (2.) we must rewrite these products of spherical harmonics as sums. Using formula \ref{eq:prod-sum} one gets:
 \begin{eqnarray}
T_{-1}^{(1)}T_{-1}^{(1)}&=&T_{-2}^{(2)}\\
T_{1}^{(1)}T_{-1}^{(1)}&=&
\frac{T_{0}^{(2)}}{\sqrt{6}}+ \frac{T_{0}^{(1)}}{\sqrt{2}} +\frac{T_{0}^{(0)}}{\sqrt{3}}\\
T_{-1}^{(1)}T_{0}^{(1)}&=&\frac{1}{\sqrt{2}}(T_{1}^{(2)}-T_{1}^{(1)})\\
T_{1}^{(1)}T_{0}^{(1)}&=&\frac{1}{\sqrt{2}}(T_{-1}^{(2)}+T_{-1}^{(1)})
\end{eqnarray}

The last two steps (3.) and (4.) are rather straightforward and give very lengthy expressions so their explicit calculation will not be shown. Mathematica\textsuperscript{\textregistered} code was developed for this two last steps and is available upon request. The final expression obtained for the electric quadrupole interaction and the the beam with OAM=1 are show in Table~\ref{tab:elems}.

\bibliographystyle{unsrt}
\bibliography{references}

\end{document}